\documentclass[aps,pre,amsmath,amssymb,reprint,superscriptaddress,floatfix,longbibliography]{revtex4-2}

\usepackage{graphicx}
\usepackage{physics}

\usepackage{dcolumn}
\usepackage{bm}
\usepackage[colorlinks=true,linkcolor=blue,citecolor=blue,urlcolor=blue]{hyperref}
\usepackage{amsmath,graphicx}
\usepackage{color}

\newcommand{\ilm}{Univ Lyon, Univ Claude Bernard Lyon 1, CNRS, Institut Lumi\`ere Mati\`ere, F-69622, VILLEURBANNE, France}

\begin{document}

\title{Complex coupling between surface charge and thermo-osmotic phenomena}

\author{Mehdi Ouadfel}
\affiliation{\ilm}
\author{Michael De San F\'eliciano}
\affiliation{\ilm}
\author{Cecilia Herrero}
\altaffiliation[Currently at ]{Laboratoire Charles Coulomb (L2C), Universit\'e de Montpellier, CNRS, 34095 Montpellier, France}
\affiliation{\ilm}
\author{Samy Merabia}
\affiliation{\ilm}
\author{Laurent Joly}
\email{laurent.joly@univ-lyon1.fr}
\affiliation{\ilm}

\date{\today}

\begin{abstract}
Thermo-osmotic flows, generated at liquid-solid interfaces by thermal gradients, can be used to produce electric currents from waste heat on charged surfaces. The two key parameters controlling the thermo-osmotic current are the surface charge and the interfacial enthalpy excess due to liquid-solid interactions. While it has been shown that the contribution from water to the enthalpy excess can be crucial, how this contribution is affected by surface charge remained to be understood. Here, we start by discussing how thermo-osmotic flows and induced electric currents are related to the interfacial enthalpy excess. We then use molecular dynamics simulations to investigate the impact of surface charge on the interfacial enthalpy excess, for different distributions of the surface charge, and two different wetting conditions. We observe that surface charge has a strong impact on enthalpy excess, and that the dependence of enthalpy excess on surface charge depends largely on its distribution. In contrast, wetting has a very small impact on the charge-enthalpy coupling. We rationalize the results with simple analytical models, and explore their consequences for thermo-osmotic phenomena. Overall, this work provides guidelines to search for systems providing optimal waste heat recovery performance.
\end{abstract}

\maketitle

\section{Introduction}
Nanofluidic systems (natural porous materials and synthetic devices where liquids are confined at the nanoscale) offer great promises to address societal challenges related to water and to energy harvesting \cite{Schoch2008,Bocquet2010,Kavokine2021}. Liquid-solid interfaces play a critical role in such nanoscale systems, and surface effects provide efficient means to produce electricity from various thermodynamic gradients available in nature. For instance, diffusio-osmotic flows, generated at liquid-solid interfaces under a gradient of salt concentration, can be used to produce electricity from the salinity difference between sea and river water 
\cite{Siria2017,Marbach2019,Joly2021}. Indeed, if the solid surface is charged, 
ions in the liquid reorganize to form a diffuse layer with an opposite charge in the vicinity of the surface, the electrical double layer (EDL) \cite{Grahame1947,IsraelachviliBook,Hartkamp2018}. The advection of the EDL by the osmotic flow then generates an electric current \cite{Fair1971,Siria2013,Mouterde2018,herrero2022chapter}.  

Similarly, thermal gradients can generate electric currents in liquids through a variety of mechanisms \cite{Bonetti2011,Bonetti2015,Dietzel2016,DiLecce2018,Zhao2021}. Among these mechanisms, 
thermo-osmotic flows \cite{Wurger2010,Barragan2017} induced by thermal gradients could be used to produce electricity from low-grade waste heat, by advecting the charge of the EDL \cite{Fu2019}. The resulting electric current is controlled by the surface charge, which is opposite to the charge in the EDL, and by the velocity of the thermo-osmotic flow. 
There is an increasing effort to better understand thermo-osmotic flows, through experimental characterization \cite{Bregulla2016,Franzl2022} and modeling \cite{ref:Ganti2017,Fu2017,Dietzel2017,Ganti2018,Fu2018,Proesmans2019,Anzini2019,Arango-Restrepo2020,chen_thermo-osmosis_2021,Anzini2022,oyarzua_water_2023,chen_thermo2023}. As detailed in the Theory section, 
for liquids, Derjaguin and collaborators developed a standard theoretical framework \cite{Derjaguin1941,derjaguin1987surface,Anderson1989,herrero2022chapter}, which relates  
the thermo-osmotic flow velocity to the interfacial enthalpy excess, stemming from the interactions of the liquid with the solid.   
This framework has been extended recently to take into account liquid-solid slippage arising on low-friction surfaces \cite{Fu2017,ref:Herrero2022,herrero2022chapter}, which can boost the flow. It has also been shown recently that, in addition to the commonly considered ion electrostatic contribution to the enthalpy excess \cite{Derjaguin1941,derjaguin1987surface,Wurger2010}, the contribution of water to the enthalpy excess could also be significant, and even dominate over the electrostatic one \cite{Fu2019,ref:Herrero2022}. While the electrostatic contribution is well described by the Poisson-Boltzmann framework \cite{Markovich2016a,Herrero_PB2,Blossey2023} (especially at low salt concentrations, at which this contribution becomes large), the water contribution to the enthalpy excess results from specific interactions with the surface and requires descriptions at the molecular level, for instance with molecular dynamics simulations. 

However, previous studies have only computed the water enthalpy excess on charge neutral surfaces \cite{ref:Herrero2022}. With the ultimate goal to use thermo-osmosis to produce electricity, it is crucial to understand how surface charge modifies the interfacial enthalpy excess, and in particular its water contribution. 
In this article, we start by clarifying the link between the interfacial enthalpy excess and thermo-osmotic phenomena. We then 
present the results of molecular dynamics simulations of an aqueous electrolyte confined between parallel charged walls. 
We investigated the impact of surface charge density on the interfacial enthalpy excess, for different distributions of the surface charge, and two different wetting conditions. We rationalized the results with simple analytical models, which can be used to evaluate the interfacial enthalpy excess in a wide variety of systems, and we explored their consequences for thermo-osmotic flows and thermo-osmotic currents.

\section{Theory}\label{sec:theory}
The purpose of this section is to show how the enthalpy excess is related to the thermo-osmotic response. 
To that aim, we will briefly recall how the standard theoretical framework initially introduced by Derjaguin and collaborators \cite{Derjaguin1941,derjaguin1987surface,Anderson1989,herrero2022chapter} can be extended to take into account liquid-solid slip \cite{Fu2017,ref:Herrero2022,herrero2022chapter}. Indeed, at the nanoscale, it is known that the standard no-slip boundary condition can fail \cite{Bocquet2007}. The velocity jump at the liquid-solid interface is quantified by the slip length $b$, which is the distance inside the wall where the linear extrapolation of the liquid velocity profile reaches the wall velocity \cite{Bocquet2007}. 

We first discuss the thermo-osmotic coefficient, which quantifies the thermo-osmotic response of a liquid-solid interface to a temperature gradient parallel to the wall, and is defined by:
\begin{equation}
    M_\mathrm{to} = -\frac{v_\mathrm{to}(\infty)}{\nabla T/T},
\end{equation}
with $T$ the temperature, and $v_\mathrm{to}(\infty)$ the thermo-osmotic velocity far from the surface; $M_\mathrm{to}$ can be positive or negative, depending on the direction of the thermo-osmotic flow. The thermo-osmotic velocity profile $v_\mathrm{to}(z)$ can be obtained by integrating Stokes equation, assuming a homogeneous viscosity $\eta$ (we will come back to this choice later) and taking into account slippage \cite{herrero2022chapter}:
\begin{equation}\label{eq:v_to}
\begin{aligned}
    v_\mathrm{to}(z) = -\frac{\nabla T/T}{\eta}\Bigg\{ & \int_0^z\dd z'\int_{z'}^\infty\delta h(z'')\dd z''\Bigg. \\
    &+ \Bigg.b\int_0^\infty\delta h(z)\dd z\Bigg\},
\end{aligned}
\end{equation}
where $z$ is the distance to the wall, $b$ is the slip length, and $\delta h(z)$ is the enthalpy excess density due to interactions between the fluid and the solid; we will discuss the definition of $\delta h$ for an electrolyte solution in the next section. 
The value of the velocity far from the surface is:
\begin{equation}
    v_\mathrm{to}(\infty) = -\frac{\nabla T/T}{\eta}\int_0^\infty(z+b)\delta h(z)\dd z,
\end{equation}
and the thermo-osmotic coefficient is thus expressed as:
\begin{equation}\label{eq:mto_coeff}
    M_\mathrm{to} = \frac{1}{\eta} \int_0^\infty (z+b)\delta h(z)\dd z.
\end{equation}
Defining the interfacial enthalpy excess (per unit surface):
\begin{equation}
    \Delta H = \int_0^\infty\delta h(z)\dd z,
\end{equation}
one can rewrite Eq.~\eqref{eq:mto_coeff} as follows:
\begin{align}
    M_\mathrm{to} &= \frac{1}{\eta}\left\{\int_0^\infty z\delta h(z)\dd z + b\Delta H \right\}\\
    &= \frac{\Delta H}{\eta}(\lambda_\mathrm{h} + b),\label{eq:thermoosm}
\end{align}
where we have set
\begin{equation}
    \lambda_\mathrm{h} = \frac{1}{\Delta H} \int_0^\infty z\delta h(z)\dd z ;
\end{equation}
$\lambda_\mathrm{h}$ is the characteristic thickness of the layer where the liquid interacts with the wall, and hence $\delta h(z) \neq 0$. For pure water, $\lambda_\mathrm{h}$ has been found to be on the order of 7\,\AA{} in previous work \cite{chen_thermo-osmosis_2021}. Generally, $\lambda_\mathrm{h}$ is controlled by the range of liquid-solid interactions, therefore its value should be similar for all water-solid interfaces.  
Note that $\lambda_\mathrm{h}$ is also the distance from the wall over which the thermo-osmotic velocity profile develops and converges to $v_\mathrm{to}(\infty)$. 

The approach developed above for thermo-osmosis is analogous to the one used to describe electro-osmotic flows, which are generated when an electric field $E$ parallel to the interface is applied. In this case, the response coefficient is given by \cite{Huang2007,Huang2008,herrero2022chapter}:
\begin{equation}\label{eq:Meo}
    M_\mathrm{eo} = \frac{v_\mathrm{eo}(\infty)}{E} = \frac{1}{\eta}\int_0^\infty (z+b)\rho_\mathrm{e}(z)\dd z,
\end{equation}
where $v_\mathrm{eo}(\infty)$ is the electro-osmotic velocity far from the surface and $\rho_\mathrm{e}$ is the charge density. To ensure electroneutrality, $\int_0^\infty\rho_\mathrm{e}(z)\dd z = -\Sigma$ with $\Sigma$ the surface charge density of the wall. Equation~\eqref{eq:Meo} can then be rewritten:
\begin{equation}\label{eq:Meo_scaling}
    M_\mathrm{eo} = -\frac{\Sigma}{\eta}(\lambda_\mathrm{eff} + b),
\end{equation}
where $\lambda_\mathrm{eff} = 1/(-\Sigma) \int_0^\infty z \rho_\mathrm{e}(z)\dd z$ is the effective Debye length \cite{Joly2006,Herrero_PB2,herrero2022chapter}, which quantifies the thickness of the EDL in the non-linear Poisson-Boltzmann regime. 
Therefore, in the same way that one can predict the electro-osmotic flow based on the surface charge density using Eq.~\eqref{eq:Meo_scaling}, one can predict the thermo-osmotic coefficient from the enthalpy excess using Eq.~\eqref{eq:thermoosm}.

Thermo-osmosis can also be used to create an electric current in the case of electrically charged surfaces 
\cite{Fu2019}. Indeed, in this case, the thermo-osmotic flow sets in motion the fluid and thus the EDL, which creates an electric current. One can quantify the thermo-osmotic current generated by thermo-osmosis for a planar surface surface of transverse width $w$ by defining the thermo-osmotic conductance $K_\mathrm{to}$: 
\begin{equation}
    K_\mathrm{to}=\frac{J_\mathrm{e}/w}{(-\nabla T/T)}=\frac{1}{-\nabla T/T}\int_0^\infty\rho_\mathrm{e}(z)\,v_\mathrm{to}(z)\dd z, 
\end{equation}
with $J_\mathrm{e}$ the thermo-electric current.
When the enthalpy excess density decreases rapidly compared to the electric potential, i.e. $\lambda_\mathrm{h} \ll \lambda_\mathrm{eff}$, 
one can consider that the thermo-osmotic velocity profile has reached its plateau value, $v_\mathrm{to}(z)\approx v_\mathrm{to}(\infty)$, everywhere in the EDL. 
The effective Debye length is a function of the ion concentration $n_0$ and the surface charge density $\Sigma$ \cite{Herrero_PB2}, so that $\lambda_\mathrm{h} \ll \lambda_\mathrm{eff}$ is met when $n_0<0.1$\,M and $\Sigma< 50$\,mC/m$^2$. In that case, $K_\mathrm{to}$ can be re-expressed as: 
\begin{align}
    K_\mathrm{to}&\approx\frac{v_\mathrm{to}(\infty)}{-\nabla T/T} \int_0^\infty \rho_\mathrm{e}(z) \dd z \approx M_\mathrm{to} \times (-\Sigma)\\
    &\approx -\frac{\Sigma\Delta H}{\eta}(\lambda_\mathrm{h}+b); \label{eq:thermoelec2}
\end{align}
$K_\mathrm{to}$ is therefore, in this limit, $M_\mathrm{to}$ scaled by the surface charge density.

Finally, let us return to the assumption of a uniform viscosity made to obtain equations \eqref{eq:thermoosm} and \eqref{eq:thermoelec2}. 
Indeed, it has been shown that, near the wall, the liquid viscosity could significantly increase \cite{ref:Ganti2017,ref:li2007,ref:bonthuis2013}. 
A good approximation for the viscosity is to consider the following step function \cite{ref:bonthuis2013}:
\begin{equation}\label{eq:viscosity}
    \eta(z) = \left\{
        \begin{array}{@{}l@{\thinspace}l}
            \xi\,\eta, & \quad  z < z_s, \\
            \eta, & \quad  z \ge z_s, 
        \end{array}
   \right.
\end{equation}
with $z_s$ the position of the plane of shear \cite{ref:bonthuis2013} and $\xi \ge 1$. Using Eq.~\eqref{eq:viscosity}, one can show that the thermo-osmotic velocity far from the surface is given by (derivation provided in the appendix):
\begin{equation}\label{eq:two_viscosity_vel}
    \begin{aligned}
        v_\mathrm{to}(\infty) = &-\frac{\nabla T/T}{\xi\,\eta}\int_0^{z_s}(z+b)\delta h(z)\dd z\\
        &- \frac{\nabla T/T}{\eta}\int_{z_s}^\infty(z+b)\delta h(z)\dd z,
    \end{aligned}
\end{equation}
and the thermo-osmotic coefficient becomes:
\begin{equation}\label{eq:two_viscosity}
    \begin{aligned}
        M_\mathrm{to} = &\frac{1}{\xi\,\eta}\int_0^{z_s}(z+b)\delta h(z)\dd z\\
        &+ \frac{1}{\eta}\int_{z_s}^\infty(z+b)\delta h(z)\dd z.
    \end{aligned}
\end{equation}
As 
expected, the thermo-osmotic coefficient is lower when considering a liquid layer near the wall with higher viscosity. The thermo-osmotic coefficient is weighted by the ratio between the two viscosities $\xi$, but still remains a function of the enthalpy excess density. In the case of a stagnant liquid layer, $\xi \to \infty$, with $b=0$, and one obtains:
\begin{equation}
    M_\mathrm{to}^{\xi\to\infty} = \frac{1}{\eta}\int_{z_s}^\infty z\,\delta h(z)\dd z<\frac{\Delta H}{\eta}\lambda_\mathrm{h},
\end{equation}
where $z_s$ is now the thickness of the stagnant layer. In this case, the thermo-osmotic coefficient is significantly lower than for a uniform viscosity, since the slip length is zero and the enthalpy excess from the stagnant layer does not contribute to the osmotic flow. On the other hand, for hydrophobic surfaces, the viscosity remains constant, even near the interface \cite{ref:li2007,ref:bonthuis2013}, $\xi = 1$, and one recovers Eq.~\eqref{eq:thermoosm}. Using hydrophobic surfaces seems therefore to be the optimal approach for maximizing osmotic responses.

Overall, Eqs. \eqref{eq:thermoosm} and \eqref{eq:thermoelec2} highlight the key role of enthalpy excess and slip length in thermo-osmotic responses. 
Accordingly, to predict thermo-osmotic responses on charged surfaces, it is crucial to know how these two quantities depend on the surface charge. While the surface charge dependence of the slip length has been investigated before  \cite{Xie2020,Mangaud2021}, less is known about the enthalpy excess, and in particular about its water contribution. In the following, we will use molecular dynamics simulations to compute $\Delta H$ as a function of the surface charge.

\section{Simulation methods}

\subsection{System}

\begin{figure}
    \centering
    \includegraphics[width=0.9\columnwidth]{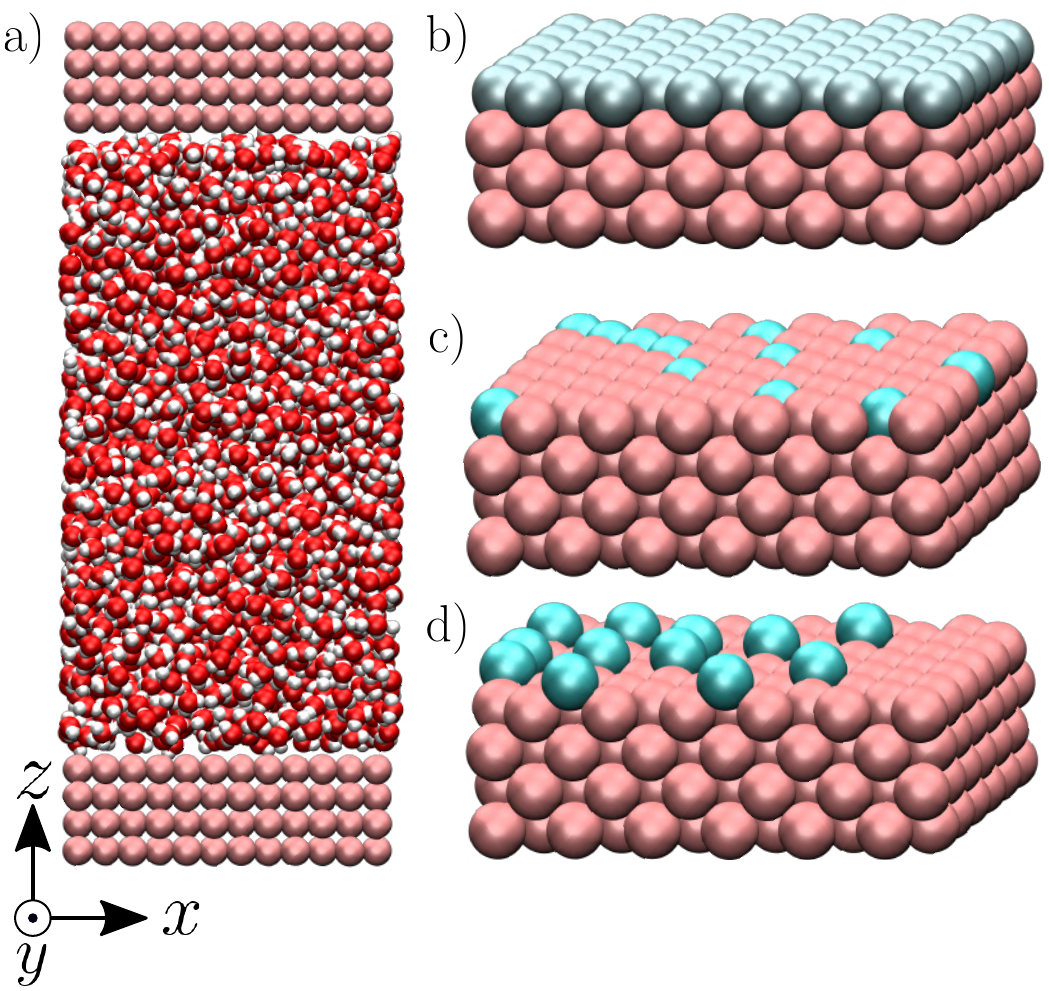}
    \caption{Visualization of the modeled systems composed of an aqueous electrolyte solution confined between two walls (a), realized with VMD \cite{ref:VMD}.\
    The walls are charged either homogeneously (b), heterogeneously (c) or with the charge protruding from the walls to create a ``defective solid'' case (d)\
        (see text for more details).}
    \label{fig:system}
\end{figure}

We used the LAMMPS package \cite{ref:LAMMPS} to perform equilibrium molecular dynamics simulations of an aqueous electrolyte composed of 2000 water molecules and NaCl salt with a bulk concentration $n_0 \sim 0.20$\,M, corresponding to a Debye length $\lambda_\mathrm{D} \sim 7$\,\AA{}, confined between two parallel walls made of four atomic layers of a fcc crystal with a lattice parameter $a=5.3496$\,\AA{} (Fig.~\ref{fig:system}). 
We used a large salt concentration, so that the ion contribution is negligible as compared to the one of water \cite{ref:Herrero2022}. 
We applied periodic boundary conditions along the $x$ and $y$ directions to our system of size $L_x = L_y = 32.0976$\,\AA{}. Water molecules were simulated using the SPC/E model \cite{ref:Berendsen1987}, which employs both Lennard-Jones (LJ)  and Coulombic potentials to model the atomic interactions. The LJ potential is defined by the characteristic diameter $\sigma_{ii}$, and the interaction particle $\varepsilon_{ii}$ of particle $i$.
For the ions, we used the LJ parameters given in Ref.~\citenum{ref:Koneshan1998} along with the Lorentz-Berthelot mixing rules. As for the LJ wall, we chose the parameters to build either a hydrophobic or a hydrophilic surface, following Ref.~\citenum{Huang2007}. We studied three types of surface charge distribution: 
1) A homogeneous case where we charged all surface atoms with a charge $q = \Sigma\,S / N_\mathrm{wall}$, where $S = L_x L_y$ is the surface of the wall and $N_\mathrm{wall}$ the number of atoms on the surface, which results in a surface charge density $\Sigma$ (Fig.~\ref{fig:system}.b); 
2) A heterogeneous case where we randomly selected atoms of the surface and attributed them a charge of $\pm$1\,e (Fig.~\ref{fig:system}.c) so that the surface charge density was $\Sigma$; 
3) A case in which atoms with a charge of $\pm$1 e protrude from the surface with respect to the fcc structure of the crystal, simulating defects on the surface (Fig.~\ref{fig:system}.d). 
In all cases, counter-ions were added to the system to keep it electrically neutral. The bottom wall was frozen and we used the top wall as a rigid piston during an equilibration phase that lasted 0.6\,ns, before fixing it at its equilibrium position to set the pressure to 10\,atm, following previous studies \cite{Huang2007, ref:Herrero2022}. The equilibrium distance between the walls was $d\sim 60$\,\AA{}. We fixed the temperature of the fluid at 298\,K via a Nosé-Hoover thermostat with a damping time of 100\,fs. The simulations lasted 10\,ns with a timestep of 2\,fs.

\subsection{Quantities of interest}\label{sec:quantities}
For a mixture of particles, we define the enthalpy excess density as \cite{ref:Ganti2017}:
\begin{equation}\label{eq:enthlapy_excess}
    \delta h(z) = \sum_i  n_i(z)[h_i(z) - h_i^\mathrm{B}],
\end{equation}
with $i\in[\mathrm{O},\mathrm{H},\mathrm{Na},\mathrm{Cl}]$ the atom type, $n_i$ the number density, $h_i$ the enthalpy per particle, and the superscript B denotes a bulk quantity, i.e. its value far from the surface, where it is homogeneous. We define the enthalpy per particle as:
\begin{equation}\label{eq:enthalpy_per_particle}
    h_i(z) = u_i(z) + \dfrac{p^{\parallel}(z)}{n_\mathrm{tot}(z)},
\end{equation}
where $u_i$ is the internal energy per particle, $n_\mathrm{tot}(z) = \sum_i n_i(z)$ the total number density and $p^{\parallel}(z)$ the components of the virial pressure tensor parallel to the surface ($p^{xx}$ or $p^{yy}$, which are equal). Indeed, pressure is anisotropic near the wall, and following previous studies \cite{ref:Ganti2017, Anzini2022}, we consider the pressure component parallel to the surface to compute the enthalpy excess density. We can consider only the contribution of the potential energy to calculate the internal energy term. Indeed the equipartition theorem implies that the kinetic energy terms cancel out: $\langle u_\mathrm{k}(z)\rangle = u_\mathrm{k}^\mathrm{B}$. Here attributing $p^{\parallel}(z)/n_\mathrm{tot}(z)$ to each atom amounts to evenly distribute the atomic volume regardless of the atom type.

From the definition of the enthalpy excess density, Eq.~\eqref{eq:enthlapy_excess}, $\delta h(z)=0$ in the absence of particles, and the enthalpy per particle needs to be defined only when the density is non zero.  
Denoting $z_0$ the minimum height at which fluid particles are found, Eq.~\eqref{eq:enthlapy_excess} becomes:
\begin{equation} \label{eq1}
\delta h(z) = 
\begin{cases}
\sum_i n_i(z)[u_{\mathrm{p},i}(z)-u_{\mathrm{p},i}^{\mathrm{B}}] & \\
\quad \quad + p^{\parallel}(z) - \dfrac{p^\mathrm{B}}{n_\mathrm{tot}^\mathrm{B}}n_\mathrm{tot}(z) & \text{if } z \ge z_0 , \\
0 & \text{if } z < z_0 , 
\end{cases}
\end{equation}
where we denote $p^\mathrm{B}$ the bulk pressure, which is isotropic, and where $u_{\mathrm{p},i}$ is the potential energy of particle type $i$, which takes into account its interactions with all atom types. 

We compute pressure profiles using the stress per atom approach. Indeed the virial part of the stress per atom is given by $\Pi^{\alpha\beta}_i = - \sum^{N_i}_k r_{k\alpha} f_{k\beta}$ where $\alpha$ and $\beta \in \{x, y, z\}$, and $N_i$ is the number of atoms of type $i$. With this definition, $p_{\alpha \beta}(z) = -(1/V)\sum_i \Pi^{\alpha \beta}_i(z)$. Although it is well known that the pressure tensor is not uniquely defined for an inhomogeneous fluid near an interface \cite{ref:shi_perspective_2023}, 
this is not an issue here since we will consider its integral, Eq.~\eqref{eq:Ptot}, which is unambiguously defined \cite{ref:shi_perspective_2023}.

Finally, we can compute the enthalpy excess by integrating the enthalpy excess density from $z_0$ to the middle of the channel: $\Delta H = \int_{z_0}^{h/2} \delta h(z) \, \dd z$, which we decompose into three contributions: the water internal energy excess $\Delta U_{\mathrm{water}}$, the ions internal energy excess $\Delta U_{\mathrm{ions}}$ and a pressure excess term $\Delta P^\parallel$:
\begin{equation}\label{eq:Uwater}
\begin{aligned}
    \Delta U_{\mathrm{water}} = \int_{z_0}^{h/2}& \{n_{\mathrm{O}}(z)[u_{\mathrm{p,O}}(z)-u_{\mathrm{p,O}}^{\mathrm{B}}]\\
    &+n_\mathrm{H}(z)[u_{\mathrm{p,H}}(z)-u_{\mathrm{p,H}}^{\mathrm{B}}]\}\,\dd z,
\end{aligned}
\end{equation}
\begin{equation}\label{eq:Uions}
\begin{aligned}
    \Delta U_{\mathrm{ions}} = \int_{z_0}^{h/2}& \{n_{\mathrm{Na}}(z)[u_{\mathrm{p,Na}}(z)-u_{\mathrm{p,Na}}^{\mathrm{B}}]\\
    &+n_\mathrm{Cl}(z)[u_{\mathrm{p,Cl}}(z)-u_{\mathrm{p,Cl}}^{\mathrm{B}}]\}\,\dd z,
\end{aligned}
\end{equation}
\begin{equation}\label{eq:Ptot}
    \Delta P^\parallel = \int_{z_0}^{h/2} \left[p^\parallel(z) - \dfrac{p^\mathrm{B}}{n_\mathrm{tot}^\mathrm{B}}n_\mathrm{tot}(z)\right]\dd z.
\end{equation}

To estimate the thermo-osmotic and thermoelectric responses of our systems, we also computed the slip length $b$ using non-equilibrium molecular dynamics simulations. With that regard, we moved the walls in opposite $x$ directions with a speed $V_x \in [10,40]\,\mathrm{m/s}$ (we verified that we were in the linear response regime), generating a linear velocity profile far from the wall. The slip length can then be determined using the Navier boundary condition \cite{Cross2018,Fu2019}:
\begin{equation}
    b = \frac{v_s}{\dot{\gamma}},
\end{equation}
with $\dot{\gamma}$ the bulk shear rate and $v_s$ the slip velocity; $v_s$ is defined as the difference between the wall velocity and the velocity of the fluid at the hydrodynamic wall position, given by $\dot{\gamma} h/2$, where the hydrodynamic height $h$ of the liquid is given by \cite{Herrero2019}:
\begin{equation}
    h = \frac{M}{\rho_\mathrm{bulk}A},
\end{equation}
with $M$ the total mass of the fluid, $\rho_\mathrm{bulk}$ the bulk mass density and $A$ the wall surface area. 

We performed four independent simulations to determine the error on the calculation of $\Delta H$. The slip length and its error were determined from measurements that fell in the linear response regime.
The error bars given in the following figures correspond to a statistical error with 95\,\% confidence level.

\section{Results and discussion}\label{sec:results}
\subsection{Impact of surface charge density}

\begin{figure}
    \centering
    \includegraphics[width=0.9\columnwidth]{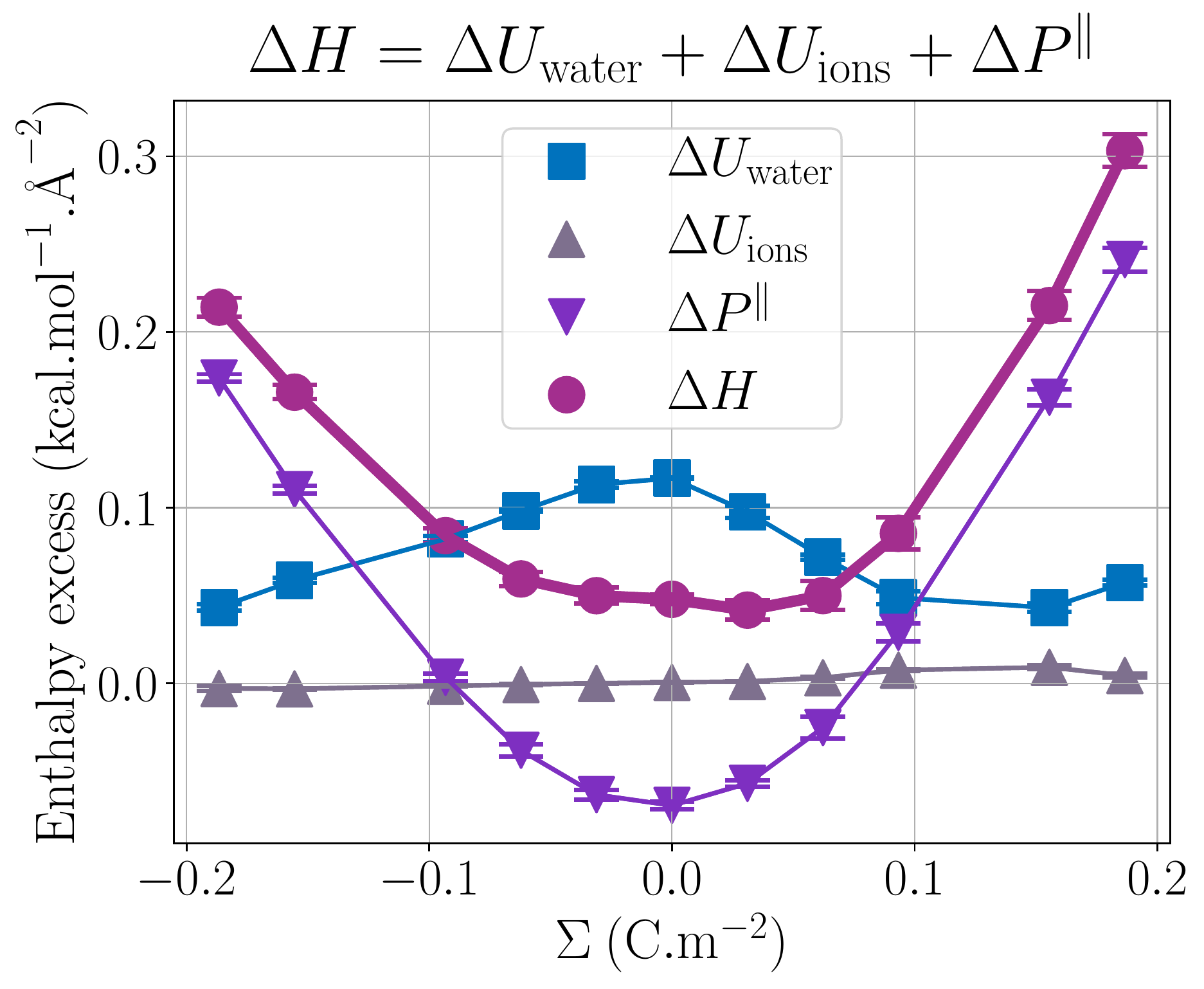}
    \caption{Comparison of the different contributions to $\Delta H$ (mauve circles): $\Delta U_{\mathrm{ions}}$ (grey triangles up) the contribution of the potential energy of the ions, $\Delta U_{\mathrm{water}}$ (blue squares) the contribution of the water potential energy and $\Delta P^\parallel$ (purple triangles down) the pressure excess.}
    \label{fig:excess_total}
\end{figure}

Figure \ref{fig:excess_total} presents the enthalpy excess $\Delta H$ and its contributions as a function of the surface charge density. One can observe that $\Delta H$ varies notably with $\Sigma$, increasing significantly with the absolute value of the surface charge density. Even at the large salt concentration considered here and at large surface charges, ions do not have much impact on the enthalpy excess, because their number remains much smaller than the number of water molecules. 
It is actually $\Delta U_{\mathrm{water}}$ and $\Delta P^\parallel$ which contribute the most to the enthalpy excess.

Let us first focus on $\Delta U_{\mathrm{water}}$. The water energy term has a parabolic form, which is relatively symmetrical with respect to the surface charge density. One can approximate this quantity using a simple dipole model. At the interface, water molecules orient themselves under the effect of the electric field created by the charged wall. In this regard the energy excess density $\delta u_{\mathrm{water}}^\mathrm{dp}$ can be expressed as:
\begin{equation}
    \delta u_{\mathrm{water}}^{\mathrm{dp}}(z) = -\langle\mu_z\rangle(z) n_\mathrm{O}(z) E(z),
\end{equation}
with $n_\mathrm{O}$ the number density of oxygen atoms, $E(z) = \Sigma/\varepsilon_0\varepsilon_\mathrm{r}(z)$ the electrostatic field where $\varepsilon_0$ is the vacuum permittivity and $\varepsilon_\mathrm{r}(z)$ are the local relative permittivity, and $\langle\mu_z(z)\rangle=\mu \langle \mathrm{cos}(\theta)(z)\rangle$ the average dipole moment along the $z$ axis, with $\langle \mathrm{cos}(\theta)\rangle$ the average dipole moment orientation and $\theta$ the angle formed by the dipole with the surface, and $\mu=1.85$\,D the water dipole moment. In the EDL, the electric field is weak with respect to its value at the surface (i.e., for $z=0$) and $\delta u^\mathrm{dp}$ is negligible \cite{ref:Herrero2022}. However, for water molecules in the first two interfacial layers, the perpendicular (or out-of-plane) relative permittivity $\varepsilon_\perp$ is greatly reduced \cite{ref:itoh2015, ref:dufils2022}. This reduction is due to the preferential orientation of water molecules close to the wall along the $z$ direction, which reduces the polarizability of water and thus the dielectric constant in this direction. The electric field becomes stronger and $\delta u^\mathrm{dp}$ is not negligible anymore.

\begin{figure}
    \includegraphics[width=1.0\columnwidth]{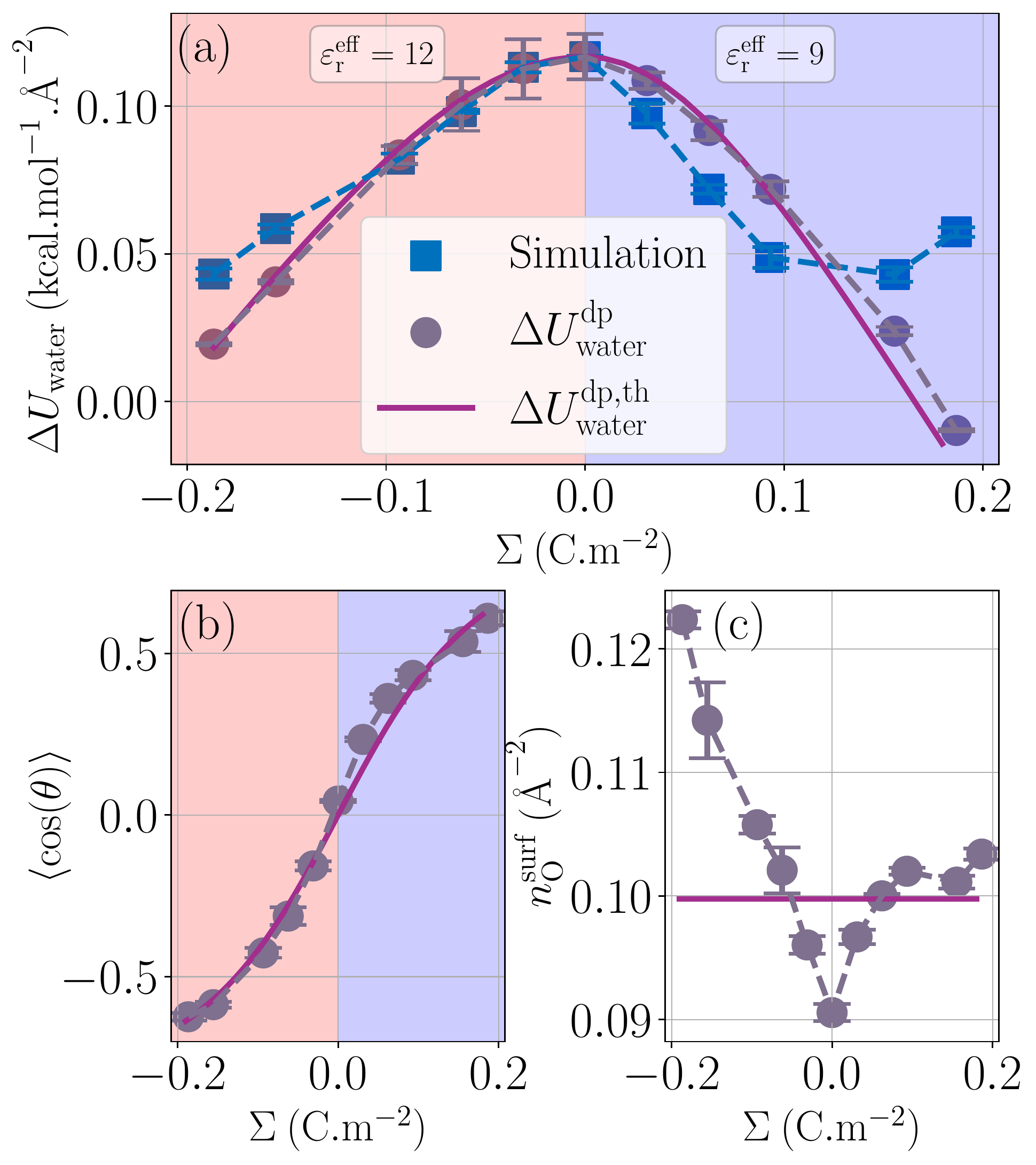}
    \caption{Water energy contribution to the enthalpy excess (a) with the measured values $\Delta U_\mathrm{water}$ (blue squares), the theoretical model with measured $\langle\mathrm{cos}(\theta)\rangle$ and $n_\mathrm{O}^\mathrm{surf}$ (grey circles), and the fully theoretical model (mauve line). The theoretical and measured orientations of water molecules near the surface (respectively mauve line and grey circles) are given in (b). The effective permittivity was set to $\varepsilon_\mathrm{r}^\mathrm{eff}=12$ for negatively charged surfaces, and $\varepsilon_\mathrm{r}^\mathrm{eff}=9$ for positively charged surfaces. The surface density of water molecules near the interface is given in (c) with the same legend as in (b).}
    \label{fig:dipole_model}
\end{figure}

We can compute $\langle \mathrm{cos}(\theta)\rangle$ from the simulation or we can compute it theoretically using Boltzmann statistics (Fig.~\ref{fig:dipole_model}.b). Let $P(\theta)$ be the probability for a molecule to have an orientation angle $\theta$, we have:
\begin{equation}
    P(\theta) = \dfrac{e^{\alpha \mathrm{cos}(\theta)}}{\int e^{\alpha \mathrm{cos}(\theta)} \dd\Omega},
\end{equation}
with $\Omega$ the solid angle and $\alpha = \beta \mu E$ with $\beta = 1/k_B T$. Thus,
\begin{equation}\label{eq:costheta_theo}
    \langle \mathrm{cos}(\theta)\rangle = \int P(\theta) \mathrm{cos}(\theta) \dd\Omega = \mathrm{coth}(\alpha) - \dfrac{1}{\alpha}.
\end{equation}
As stated before, near the interface, the relative permittivity decreases significantly and we can no longer consider the bulk value $\varepsilon_\mathrm{r} = 71$ for SPC/E water \cite{ref:REDDY1989, ref:Braun2014}. This is the reason why we considered an effective relative permittivity $\varepsilon_\mathrm{r}^{\mathrm{eff}}$, treated as a fitting parameter. Thus, writing $E=\Sigma/\varepsilon_0\varepsilon_\mathrm{r}^\mathrm{eff}$, one obtains:
\begin{equation}
    \delta u_{\mathrm{water}}^{\mathrm{dp}} = -\dfrac{\mu \left(\mathrm{coth}(\alpha) - \dfrac{1}{\alpha}\right) \Sigma}{\varepsilon_0\varepsilon_\mathrm{r}^\mathrm{eff}}n_\mathrm{O}(z),
\end{equation}
and so:
\begin{equation}
    \Delta U_{\mathrm{water}}^{\mathrm{dp}} = -\dfrac{\mu \left(\mathrm{coth}(\alpha) - \dfrac{1}{\alpha}\right) \Sigma}{\varepsilon_0\varepsilon_\mathrm{r}^{\mathrm{eff}}}n_\mathrm{O}^\mathrm{surf},
\end{equation}
with $n_\mathrm{O}^\mathrm{surf}$ the atomic surface density of the first layer, which can be computed from the simulations. Finally $\Delta U_\mathrm{water} = \Delta U_\mathrm{water}^0 + \Delta U_{\mathrm{water}}^{\mathrm{dp}}$, where $\Delta U_\mathrm{water}^0$ is the value of $\Delta U_\mathrm{water}$\
for an electrically neutral surface.

Fig.~\ref{fig:dipole_model}.b represents the dipole orientation, measured in the simulations, and fitted using Eq.~\eqref{eq:costheta_theo}. We can see an asymmetry of the measured curve, where the dipole orientation follows two different patterns: for the negative surface charges, $\theta$ precisely follows its theoretical value by taking $\varepsilon_\mathrm{r}^{\mathrm{eff}}=12$. However, for the positive ones, the model is less accurate. To fit this part of the curve, we took $\varepsilon_\mathrm{r}^{\mathrm{eff}}=9$. As mentioned above, the dielectric constant of the first layers depends on the orientation of water molecules. Its asymmetric behavior, depending on the sign of the surface charge density, is therefore expected. 

Figure~\ref{fig:dipole_model}.c presents the atomic surface density of the first layer, measured in the simulations, and 
approximated by $1/\sigma_{\mathrm{O}}^2 \simeq 0.10$\,\AA{}$^{-2}$, with $\sigma_{\mathrm{O}}$ the oxygen LJ diameter.
We thus have two ways of plotting the model presented (Fig.~\ref{fig:dipole_model}.a): from direct measurement of $\langle \mathrm{cos}(\theta)\rangle$ and $n_\mathrm{O}^\mathrm{surf}$, or from their theoretical estimates. Overall, although the asymmetry of $\mathrm{cos}(\theta)$ and $n_\mathrm{O}^\mathrm{surf}$ is not sufficient to explain the asymmetry found in $\Delta U_{\mathrm{water}}$, our model describes fairly the water excess energy term.

\begin{figure}
    \centering
    \includegraphics[width=0.9\columnwidth]{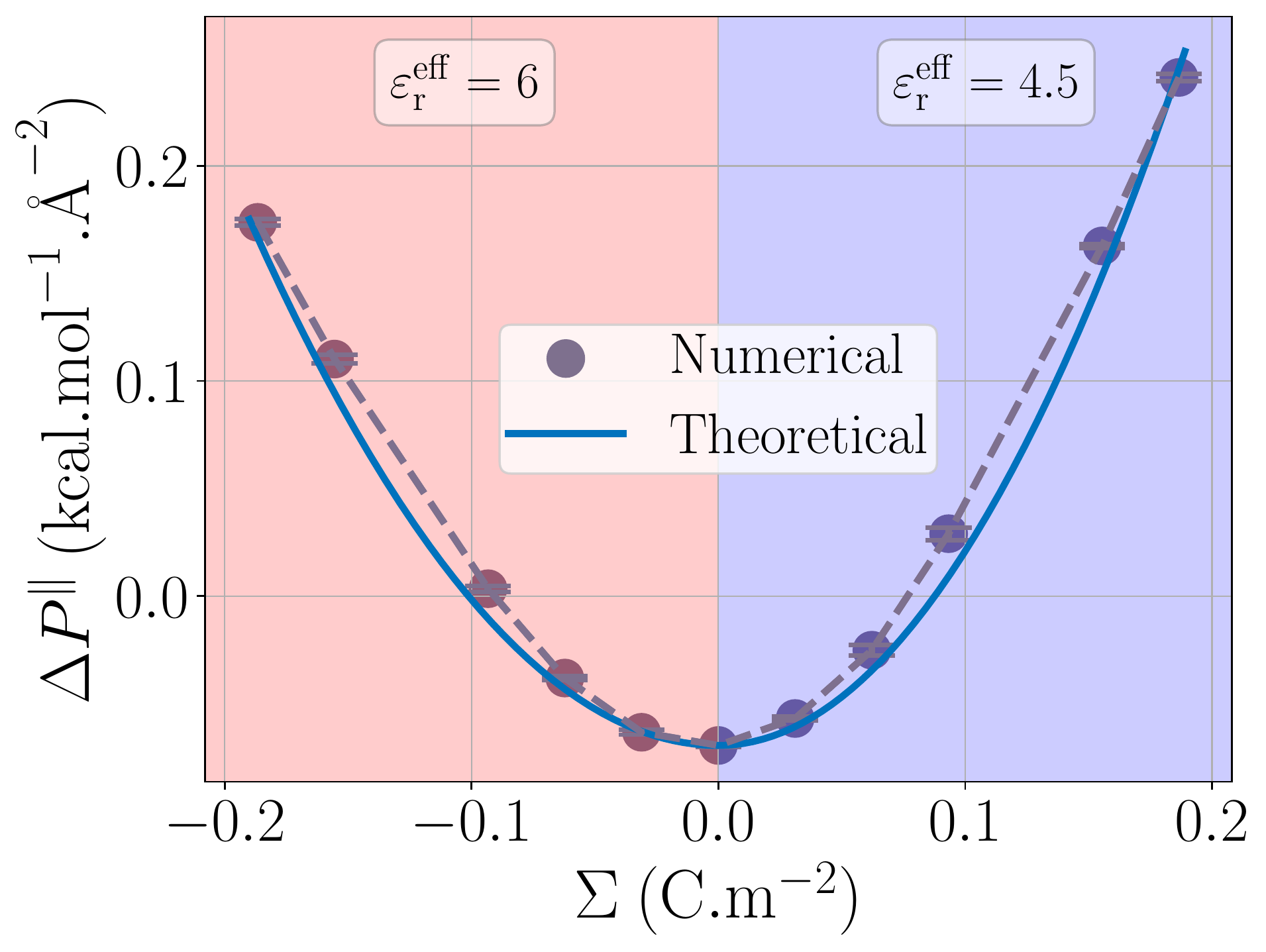}
    \caption{Pressure term $\Delta P^{\parallel}$ as a function of the surface charge density. The value obtained from the simulations (grey points) is rationalized by a capacitor model (blue line). The effective permittivity in Eq.\eqref{eq:deltaP_par} is set to $\varepsilon_\mathrm{r}^\mathrm{eff}=6$ for negatively charged surfaces, and $\varepsilon_\mathrm{r}^\mathrm{eff}=4.5$ for positively charged surfaces.}
    \label{fig:p_par_homo}
\end{figure}
Regarding the pressure term (Fig.~\ref{fig:p_par_homo}), one way to describe it is to make an analogy between $\Delta P^\parallel$ and the surface tension $\gamma$. Indeed the surface tension of a liquid-solid interface can be computed by the mechanical route \cite{dreher_calculation_2018,dreher_calculation_2019}:
\begin{align}\label{eq:gamma_mechanics}
    \gamma &= \int_{-\infty}^{+\infty} [p_{\perp}(z) - p_{\parallel}(z)] \dd z,
\end{align}
where $p_{\perp}$ and $p_{\parallel}$ are the normal and tangential components of the pressure tensor. {To ensure mechanical equilibrium,} it is necessary to have a constant perpendicular pressure along the channel, $p_\perp(z) = p_\perp^\mathrm{B} = p_{\parallel}^\mathrm{B}$, because pressure is isotropic in a bulk liquid. Thus, the surface tension becomes:
\begin{equation}
    \gamma = \int_{-\infty}^{+\infty}[p_{\parallel}^\mathrm{B}-p_{\parallel}(z)]\dd z\sim-\Delta P^\parallel.
\end{equation}
We will therefore try to describe the variation of $\Delta P^\parallel$ with $\Sigma$ following standard electrowetting models. 
The variation of the surface tension with respect to the surface charge density is given by the Lippmann's equation \cite{ref:Lippmann1875, ref:kramer2007}, which considers the energy stored in the capacitor formed by the charged surface and the EDL:
\begin{equation}\label{eq:gamma_electric}
    \gamma = \gamma_0 - \dfrac{\Sigma^2}{2C} = \gamma_0 - \dfrac{d\Sigma^2}{2\varepsilon_0\varepsilon_\mathrm{r}^\mathrm{eff}},
\end{equation}
where $\gamma_0$ is the surface tension for a neutral surface and $C = \varepsilon_0\varepsilon_\mathrm{r}^\mathrm{eff}/d$ is the capacitance per unit area, with $d$ the capacitor thickness, i.e. the mean distance between charges on the wall and counter-ions in the EDL. Similarly, one can define a capacitor like model to understand the variation of the pressure excess with the surface charge density:
\begin{equation}\label{eq:deltaP_par}
    \Delta P^\parallel = \Delta P^{\parallel, \Sigma=0} + \frac{d\Sigma^2}{2\varepsilon_0\varepsilon_\mathrm{r}^\mathrm{eff}}.
\end{equation}
Once again, the relative permittivity $\varepsilon_\mathrm{r}^\mathrm{eff}$ used in the model must be smaller than the one of bulk water due to its drop near the interface. Moreover we can see in Fig.~\ref{fig:dipole_model}.c that near the surface, water displays different structuring depending on the sign of $\Sigma$, and so should do the relative permittivity. This allows us to fit our capacitor model with two values of the relative permittivity, $\varepsilon_\mathrm{r}^\mathrm{eff} = 6$ for the negative charge surfaces $\varepsilon_\mathrm{r}^\mathrm{eff} = 4.5$ for positive ones, see Fig.~\ref{fig:p_par_homo}.

\begin{figure}
    \centering
    \includegraphics[width=1.0\columnwidth]{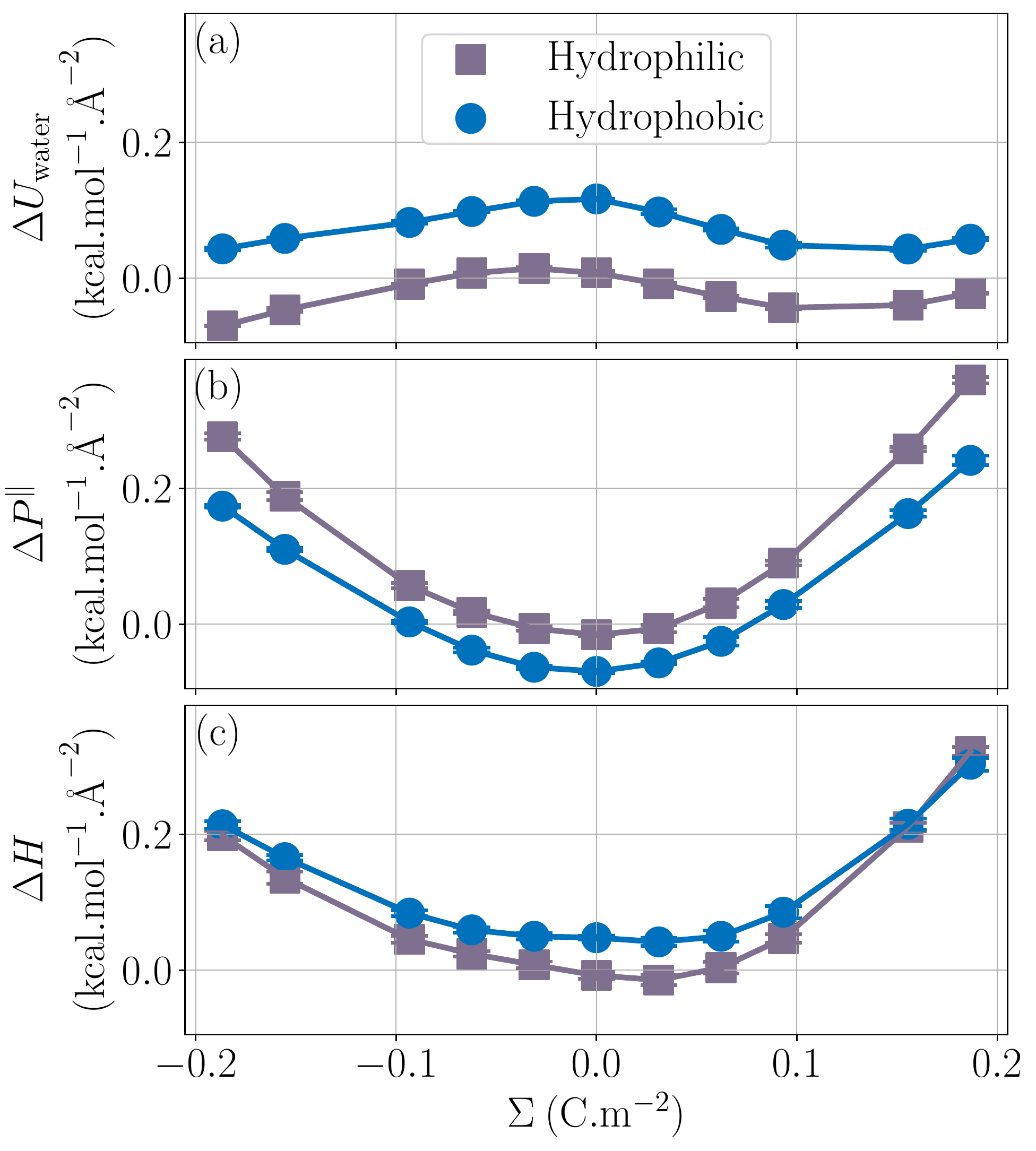}
    \caption{Impact of wetting on the enthalpy excess. Two wettings are explored here: a hydrophobic case (blue circles) and a hydrophilic case (grey squares). The water contribution (a), the pressure term (b), as well as the total enthalpy excess (c) are represented. 
    }
    \label{fig:wetting_effect}
\end{figure}

\subsection{Effect of wetting properties}
Now that we have understood the behavior of the enthalpy excess for a hydrophobic and homogeneously charged wall, let us look at the effect of wetting on the enthalpy excess. In Fig.~\ref{fig:wetting_effect}.c we compared the enthalpy excess calculated close to a hydrophobic and a hydrophilic surface. First we note that the hydrophobic surfaces have higher enthalpy excess, which is consistent with a previous study on thermo-osmosis that indicates that hydrophobic surfaces maximize the thermo-osmotic response\cite{ref:Herrero2022}. Secondly, for an electrically neutral surface, there is a change of sign of the enthalpy excess, indicating a change of direction of the thermo-osmotic flow, again being consistent with a previous study that reports a change of direction on the flow when changing the wetting of the system \cite{Fu2017}. On silica-based materials, which have hydrophilic surfaces, it has been shown that the surface charge density can affect the thermo-osmotic flow direction, this effect being attributed to a change of sign of the enthalpy excess \cite{Bregulla2016,chen_thermo2023}, our calculations tend to confirm qualitatively the results of these studies. While the enthalpy of a neutral surface depends on the wetting, its variation with the surface charge hardly depends on the wetting and one can say as a first approximation that the wetting simply shifts the values of $\Delta H$. When looking at the decomposition of the enthalpy excess, it appears that $\Delta U_\mathrm{water}$ is simply shifted and becomes smaller for hydrophilic surfaces. Water molecules are indeed more attracted to the wall in the hydrophilic case, which allows them to adopt a more favorable energy configuration thus reducing the overall internal energy (Fig.~\ref{fig:wetting_effect}.a). The pressure term is also shifted and the parabola is more pronounced. One can use the capacitor model to understand this: in the hydrophilic case, water molecules are closer to the surface, thus the capacitor is thinner, resulting in a variation of the capacitance and a modification of the curvature of $\Delta P^\parallel$. Note that we simulated two systems with very different wetting properties, and the change in capacitance remained small, so that considering that the capacitance is independent of wetting represents a good approximation.

\subsection{Effect of charge distribution on the enthalpy excess}
\begin{figure}
    \centering
    \includegraphics[width=1.0\columnwidth]{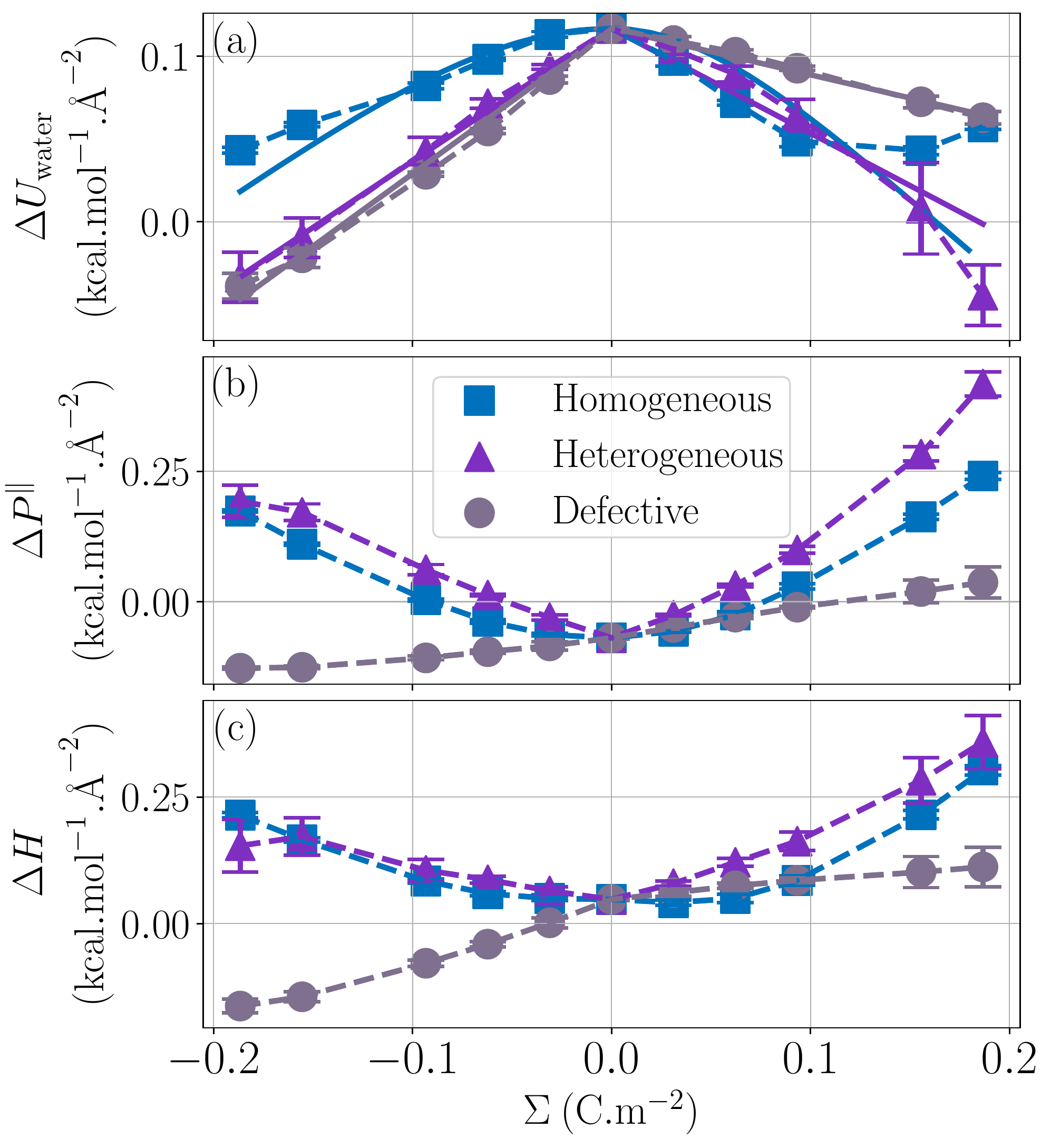}
    \caption{Effect of the charge distribution on the enthalpy excess. For $\Delta U_\mathrm{water}$ (a), the homogeneous case (blue squares) is described by a dipole model (blue line) while the heterogeneous (purple triangles) and the defective (grey circles) cases are described by a linear model of potential energy excess per unit charge, Eq.~\eqref{eq:energy_excess}, represented in full lines. For $\Delta P^\parallel$ (b), the homogeneous as well as the heterogeneous case can be described by a capacitor model but the model cannot describe the behavior of the defective solid. There is a change of sign of $\Delta H$ for the defective solid case, taking negative values for negatively charged surfaces (c).}
    \label{fig:distribution_effect}
\end{figure}
Regarding the charge distribution, one can see in Fig.~\ref{fig:distribution_effect} that it has a drastic impact on the enthalpy excess. In particular, the defective solid case displays a different behavior than the other two, displaying a change of sign for negatively charged surfaces. One can understand these differences by looking at the decomposition of $\Delta H$. For $\Delta U_\mathrm{water}$ (Fig.~\ref{fig:distribution_effect}.a), the dipole model, which assumes a homogeneous charge distribution, does not work for the heterogeneous and defective solid cases. Indeed, in the latter cases, a charge will only affect the surrounding water molecules, because with a heterogeneous charge distribution, a counter-ion will tend to bond to each charge, with the effect to screen the electric field. 
Therefore the variation of the energy excess is simply proportional to the number of surface charges:
\begin{equation}\label{eq:energy_excess}
    \Delta U_\mathrm{water} = \Delta U_\mathrm{water}^{0} + U^\mathrm{ES} \dfrac{|\Sigma|}{e}
\end{equation}
where $U^\mathrm{ES}$ is the energy excess per charge, which is a function of $\mathrm{sgn}(\Sigma)$ and the charge distribution (Table~\ref{tab:water_enthalpy_constant}), and $e$ is the elementary charge. We obtain the values of the energy excess per charge by linearly fitting the measured values of $\Delta U_\mathrm{water}$ with Eq.~\eqref{eq:energy_excess}, see Fig.~\ref{fig:distribution_effect}.

\begin{table}[h]
\centering
\begin{tabular}{@\quad l@\quad|@\quad l@\quad|@\quad l}
  $U^\mathrm{ES}$ ($\mathrm{kcal}.\mathrm{mol}^{-1}$) & $\Sigma < 0$ & $\Sigma > 0$\\
  \hline
  Heterogeneous & $-1.24\times10^{-2}$ & $-9.8\times10^{-3}$\\
  \hline
  Defective & $-1.32\times10^{-2}$ & $-4.3\times10^{-3}$\\
\end{tabular}
\caption{Water enthalpy excess per charge for different charge distributions.}
\label{tab:water_enthalpy_constant}
\end{table}

Even though there is a quantitative difference between the three cases for $\Delta U_\mathrm{water}$, the differences on the enthalpy excess come largely from $\Delta P^\parallel$ (Fig.~\ref{fig:distribution_effect}.b). For the heterogeneously charged surface, even though the charge is not evenly distributed, the capacitor model is still applicable, due to the presence of a clear gap between water molecules and the surface. However, in the presence of defects, water molecules near the surface are on the same level as charges. In this case, the thickness of the capacitor $d$ is 0, and our model predicts that the parabolic dependence of $\Delta P^\parallel$ with $\Sigma$ should vanish. We suggest that the small remaining drift of $\Delta P^\parallel$ with the surface charge, not captured by our model, originates from specific liquid-wall interactions, indirectly affected by the surface charge.
\subsection{Consequences on transport properties}
\begin{figure}
    \centering
    \includegraphics[width=1.0\columnwidth]{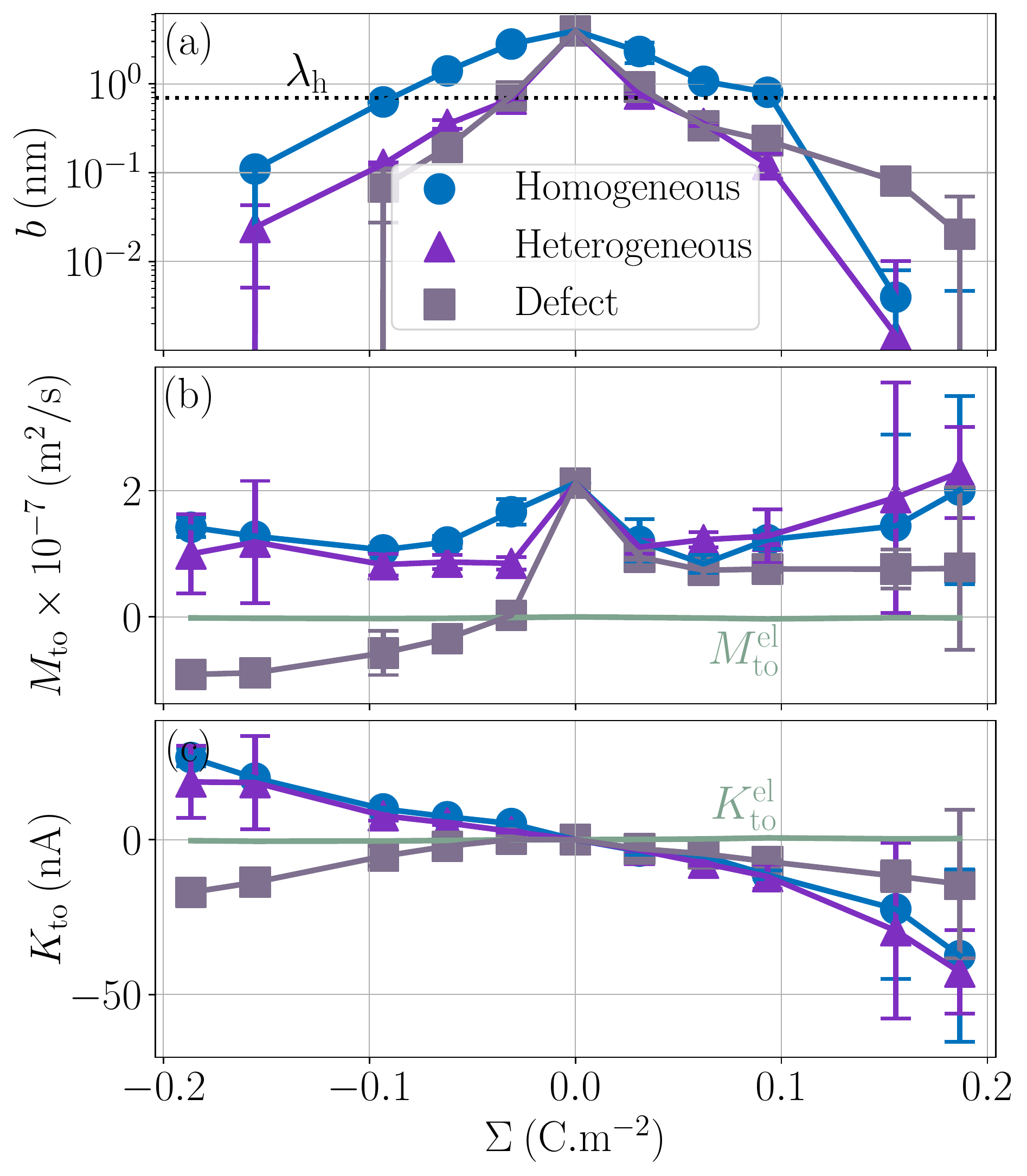}
    \caption{Slip length and resulting transport coefficients $M_\mathrm{to}$ and $K_\mathrm{to}$ as a function of the surface charge density $\Sigma$. The slip length is relatively small for this surface, and decreases rapidly with the surface charge density, in particular in the heterogeneous and defective cases; the thickness of the interaction layer $\lambda_\text{h}$ is shown with a dotted line for comparison (a). Corresponding thermo-osmotic coefficient (Eq.~\ref{eq:thermoosm}), along with the standard electrostatic prediction (Eq.~\ref{eq:th-osm_PB}) plotted in light green (b), and theoretical thermo-osmotic conductance (Eq.~\ref{eq:thermoelec2}), along with the standard electrostatic prediction (Eq.~\ref{eq:thel_PB}) plotted in light green (c).}
    \label{fig:slip_and_coefficients}
\end{figure}
Let us now look at the effect of the enthalpy excess on thermo-osmotic transport properties. To compute $M_\mathrm{to}$ and $K_\mathrm{to}$, 
we used Eqs. \eqref{eq:thermoosm} and \eqref{eq:thermoelec2}, with $\lambda_\mathrm{h} = 7$\,\AA{} and $\eta = 0.729$\,mPa$\cdot$s, the viscosity of SPC/E water at 1\,atm; it is indeed very close to the viscosity of SPC/E at 10\,atm \cite{gonzalez_shear_2010}. We plotted the results in Fig.~\ref{fig:slip_and_coefficients}.
First, one can see that the slip length decreases with the absolute value of the surface charge, but it decreases faster in both heterogeneous and defective cases \cite{Xie2020}. For the hydrophobic surface considered, the slip length is relatively small even at zero surface charge, $b^{\Sigma = 0} = 3.9$\,nm, and it decreases quickly so that $b$ becomes smaller than $\lambda_\mathrm{h}$ when the absolute surface charge density exceeds 100\,$\mathrm{mC/m}^2$ for homogeneous surfaces and 40\,$\mathrm{mC/m}^2$ in the other cases. Thus, for relatively large surface charge densities, the thermo-osmotic and thermoelectric coefficients no longer depend on $b$ and their variations with the surface charge are only driven by the enthalpy excess, together with the surface charge density for $K_\mathrm{to}$, which explains the similarity between the homogeneous and heterogeneous cases. For low surface charges, the slip length influences the response coefficients and the homogeneous charge distribution gives the best results. 

Let us see how these results compare to the classical theory, which only considers the electrostatic contribution of ions to the enthalpy excess density \cite{herrero2022chapter,ref:Herrero2022}:
\begin{equation}
    \delta h_\mathrm{el}(z)=-\varepsilon V(z) \frac{\dd^2 V}{\dd z^2} +\frac{\varepsilon}{2}\left(\frac{\dd V}{\dd z}\right)^2,
\end{equation}
where $V(z)$ is the electrostatic potential, which can be computed analytically using the Poisson-Boltzmann theory. In this case, the thermo-osmotic response and the thermo-osmotic conductance become \cite{herrero2022chapter}:
\begin{equation}\label{eq:th-osm_PB}
\begin{aligned}
    {M}_{\mathrm{to}}^{\mathrm{el}}=\frac{1}{2\pi\ell_{\mathrm{B}}\eta\beta}\biggl\{&-3\ln\bigl(1-\gamma^{2}\bigr)-\mathrm{asinh}^{2}(x)\\
    &+\frac{b}{\lambda_{\mathrm{D}}}\biggl[3x\vert\gamma\vert-2x\mathrm{asinh}(x)\biggr]\biggr\},
\end{aligned}
\end{equation}
\begin{multline}\label{eq:thel_PB}
    K_\mathrm{to}^\mathrm{el} = - \frac{e}{2\pi^2\ell_\mathrm{B}^2 \eta \beta } \frac{\mathrm{sgn}(\Sigma) x}{\lambda_{\mathrm{D}}} \Bigg\{ 5 \qty[1 - \frac{\mathrm{asinh}(x)}{x}]\\
    - 2 \abs{\gamma}\mathrm{asinh}(x) + \frac{b}{\lambda_{\mathrm{D}}}  \bigg[ 3 \abs{\gamma} x - 2 x \mathrm{asinh}(x) \bigg] \Bigg\},
\end{multline}
with $\ell_\mathrm{B} = \beta e^2/(4\pi\varepsilon)$ the Bjerrum length, where $e$ is the elementary charge, $x=\lambda_\mathrm{D}/\ell_\mathrm{GC}$, with $\lambda_\mathrm{D} = 1/\sqrt{8\pi \ell_\mathrm{B} n_0}$ the Debye length and $\ell_\mathrm{GC}=e/(2\pi\ell_\mathrm{B}\vert\Sigma\vert)$ the Gouy-Chapman length, and $\gamma=(\mathrm{sgn}(\Sigma)/x)\left\{-1+{\sqrt{1+x^{2}}}\right\}$ \cite{Herrero_PB2}.

As shown in Fig.~\ref{fig:slip_and_coefficients}, for the parameters considered in this work, i.e. $\lambda_\mathrm{D}\simeq 7$\,\AA{}, $\ell_\mathrm{B}\simeq 8$\,\AA{}, and the slip length of the homogeneous charge distribution, the classical thermo-osmotic theory predicts thermo-osmotic coefficients much lower than the ones computed with the total enthalpy excess; note however that the electrostatic contribution of ions could be larger in other range of parameters, and in particular at lower salt concentrations \cite{ref:Herrero2022}. Moreover, $\delta h_\mathrm{el}(z)$ only predicts negative thermo-osmotic coefficients $M_\mathrm{to}$, and predicts a thermo-osmotic conductance having the same sign as the surface charge density, which is also in strong contrast with the predictions taking into account the total enthalpy excess. Once again, this highlights the important contribution of the solvent to the enthalpy excess, which leads to a rich and complex behavior, such as the change of sign of the thermo-osmotic coefficient for protruding charges.

\section{Conclusion}
In this article, we highlighted the connection between the interfacial enthalpy excess and thermo-osmotic transport; in particular, we clarified the impact of the enthalpy excess on the thermo-osmotic coefficient $M_\mathrm{to}$ and the thermo-osmotic conductance $K_\mathrm{to}$. We then used equilibrium molecular dynamics simulations to study the effect of surface charge density and charge distribution on the enthalpy excess. We have shown that the surface charge density has a large impact on the enthalpy excess. For homogeneously charged surfaces, the enthalpy excess is enhanced by 300 to 500\,\% for the highest surface charge densities considered. 
We investigated the different contributions to the enthalpy excess, and showed that it was dominated by the change of tangential pressure close to the wall, with a non-negligible correction due to the change in water internal energy. In contrast, the contribution from ions internal energy was negligible for the range of parameters used in this work. 
We then rationalized how the different contributions to the enthalpy excess depended on surface charge with simple analytical models.

We also studied the effect of wetting to realize that it does not significantly modify the impact of surface charge on the enthalpy excess. Nonetheless the values of the enthalpy excess are more important in the hydrophobic case, especially at small surface charges, it is thus preferable to use non-wetting surfaces to maximize the enthalpy excess. We also studied the effect of the charge distribution on the enthalpy excess. We have shown that using homogeneously or heterogeneously charged surfaces does not have a strong impact on the enthalpy excess. However, the presence of protruding defects has a great influence on it. In particular, with this charge distribution we have been able to observe a change of sign of $\Delta H$ for negative surface charges. The strong changes in the enthalpy excess are largely reflected on the thermo-osmotic transport coefficients. The standard picture, which only considers the electrostatic contribution of ions to compute the enthalpy excess, does not capture the complexity of the thermo-osmotic responses. 
It it thus necessary to use molecular dynamics to compute the enthalpy excess close to charged surfaces. Overall we provide a useful tool to explore a wide variety of systems and identify those promising the best thermo-osmotic performance. 
For example it could be interesting to study thermo-osmotic responses of new two-dimensional materials using this method.

\begin{acknowledgments}
The authors thank D. Pandey and S. Hardt for fruitful discussions. This work is supported by the ANR, Project ANR-21-CE50-0042-01 smoothE. 
\end{acknowledgments}

\appendix

\section{Derivation of the thermo-osmotic velocity with a viscosity following a step function}\label{sec:derivation}
\begin{figure}
    \centering
    \includegraphics[width=0.9\columnwidth]{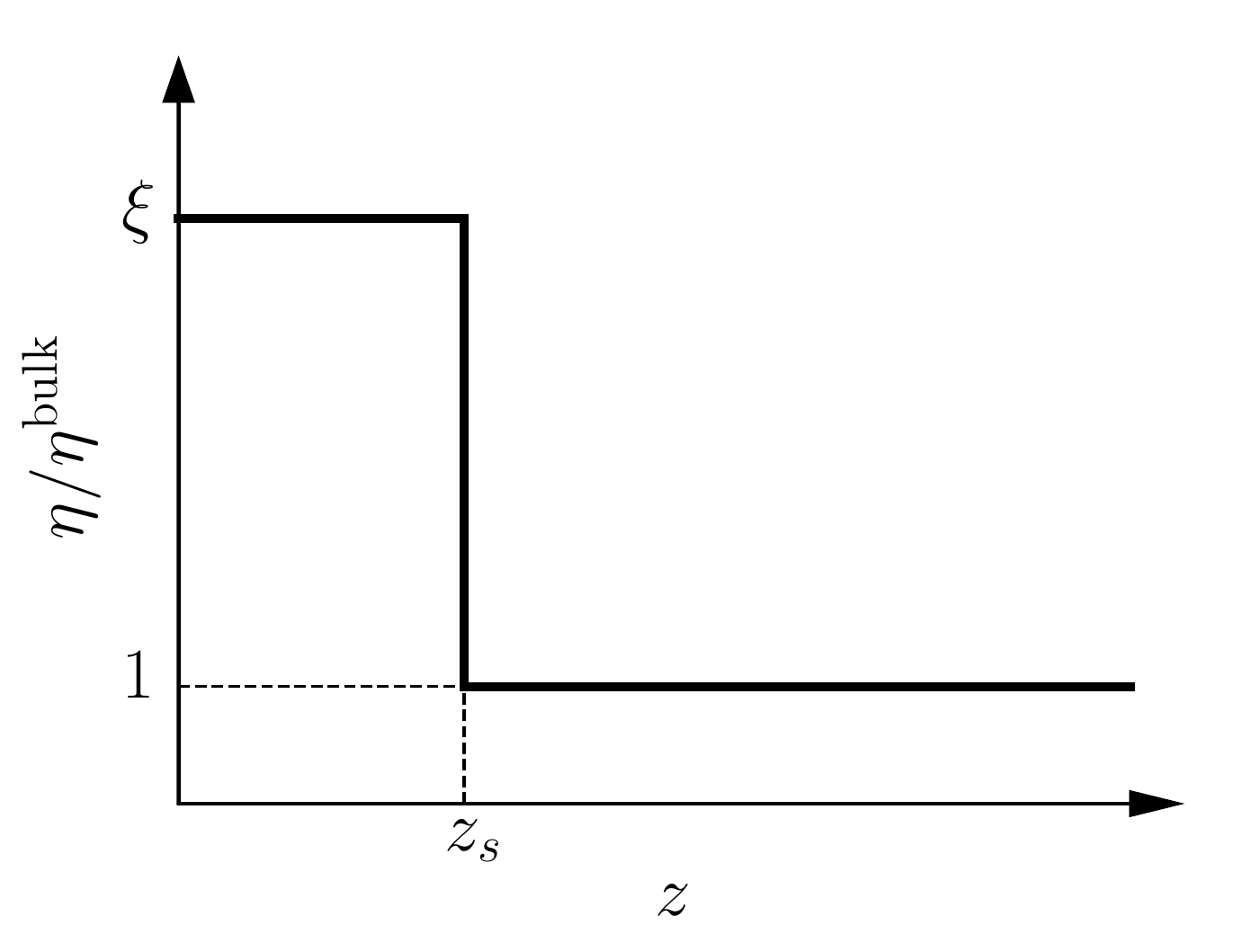}
    \caption{Viscosity as a function of the distance to the wall in the ``step'' approximation.}
\end{figure}

To derive Eq.~\eqref{eq:two_viscosity_vel}, we start from Stokes equation:
\begin{equation}
    \left\{
    \begin{aligned}
    -\eta\frac{\dd^2 v}{\dd z^2} &= f(z),\quad z>z_s\\
    -\xi\,\eta\frac{\dd^2 v}{\dd z^2} &= f(z),\quad z<z_s
    \end{aligned}
    \right.
\end{equation}
where $f(z)$ is the force density generated by a thermodynamic gradient along the interface. It is equal to $-\delta h(z)(\nabla T/T)$ for a thermal gradient. The integration of Stokes equation leads to:
\begin{align}
    \eta\frac{\dd v}{\dd z}&=\int_z^\infty f(z')\dd z',&z>z_s\label{eq:zsupzs}\\
    \xi\,\eta\frac{\dd v}{\dd z}&=\xi\int_{z_s}^\infty f(z)\dd z + \int_z^{z_s}f(z')\dd z',\quad &z<z_s.\label{eq:zinfzs}
\end{align}
We integrate these equations again. From Eq.~\eqref{eq:zsupzs} we get the velocity of the fluid far from the interface:
\begin{equation}
    v(\infty) = v(z_s) + \frac{1}{\eta}\int_{z_s}^\infty\dd z \int_{z}^\infty f(z')\dd z',
\end{equation}
which we integrate by part:
\begin{equation}
    v(\infty) = v(z_s) + \frac{1}{\eta}\int_{z_s}^\infty z\,f(z)\dd z-\frac{z_s}{\eta}\int_{z_s}^\infty f(z)\dd z.
\end{equation}
The velocity at the plane of shear $v(z_s)$ is determined from Eq.~\eqref{eq:zinfzs}:
\begin{equation}
    v(z_s)=v(0)+\frac{z_s}{\eta}\int_{z_s}^\infty f(z)\dd z + \frac{1}{\xi\,\eta}\int_0^{z_s}\dd z\int_{z}^{z_s}f(z')\dd z',
\end{equation}
which simplifies by integrating by part:
\begin{equation}
    v(z_s)=v(0)+\frac{z_s}{\eta}\int_{z_s}^\infty f(z)\dd z + \frac{1}{\xi\,\eta}\int_0^{z_s}z\,f(z)\dd z.
\end{equation}
The velocity of the fluid at the interface, $v(0)$, is given by the Navier boundary condition:
\begin{equation}
    v(0) = b\eval{\dv{v}{z}}_{z=0} = \frac{b}{\xi\,\eta}\int_0^{z_s}f(z)\dd z + \frac{b}{\eta}\int_{z_s}^\infty f(z)\dd z.
\end{equation}
Finally:
\begin{equation}
    v(\infty) = \frac{1}{\xi\,\eta}\int_0^{z_s}(z+b)f(z)\dd z + \frac{1}{\eta}\int_{z_s}^\infty(z+b)f(z)\dd z.
\end{equation}
Substituting $f(z)$ by $-\delta h(z)(\nabla T/T)$, one obtains Eq.~\eqref{eq:two_viscosity_vel}.


%


\end{document}